\documentclass[runningheads]{llncs}
\usepackage{graphicx}
\usepackage{algorithm}
\usepackage{algpseudocode}
\usepackage{color}
\usepackage{mathtools}
\usepackage{caption}
\usepackage{orcidlink}

\newcommand{\mlcomment}[1]{}

\begin{document}
\title{Online Job Failure Prediction \\ in an HPC System}
%
\author{} 
\institute{}
\author{Francesco Antici \inst{1}\orcidlink{0000-0002-1125-0588} \and Andrea Borghesi\inst{1}\orcidlink{0000-0002-2298-2944} \and Zeynep Kiziltan\inst{1}\orcidlink{0000-0003-0412-4396}}
%
%
\institute{Univeristy of Bologna, Italy \\
    \texttt{\{francesco.antici,andrea.borghesi3,zeynep.kiziltan\}@unibo.it}  
}
\maketitle              
\begin{abstract}

Modern High Performance Computing (HPC) systems are complex machines, with major impacts on economy and society. Along with their computational capability, their energy consumption is also steadily raising, representing a critical issue given the ongoing environmental and energetic crisis. Therefore, developing strategies to optimize  HPC system management has paramount importance, both to guarantee top-tier performance and to improve energy efficiency. One strategy is to act at the workload level and highlight the jobs that are most likely to fail, prior to their execution on the system. Jobs failing during their execution unnecessarily occupy resources which could delay other jobs, adversely affecting the system performance and energy consumption. 
In this paper, we study job failure prediction at submit-time using classical machine learning algorithms. Our novelty lies in (i) the combination of these algorithms with Natural Language Processing (NLP) tools to represent jobs and (ii) the design of the approach to work in an online fashion in a real system. The study is based on a dataset extracted from a production machine hosted at the HPC centre CINECA in  Italy. Experimental results show that our approach is promising. 

\end{abstract}

\section{Introduction}

High Performance Computing (HPC) is a term used in Computer Science to represent the practice of aggregating computing power to solve complex problems. HPC machines are organized in clusters and they consist of several computing units (nodes) networked together to work in parallel and boost processing speed. Nodes are connected through a low-latency internal network bus, which routes traffic to mimic the behaviour of a single computer.
The last decades have witnessed a massive increase in the number of  components, accelerators and consequently consumption of computational power of HPC centers.
This trend has been fuelled by the development of computational- hungry techniques, indeed HPC systems play a fundamental role in the field of data science, and are widely used for computationally intensive tasks in various fields, such as quantum mechanics, weather forecasting, climate research.

Latest HPC systems have reached exascale performance, namely $10^{18}$ operations per second, and in the future more systems are expected to have similar characteristics \cite{carpenter2022etp4hpc}. Machines of such scale must comply with certain standard of performance and energy efficiency, hence it is fundamental to develop strategies to optimize their workload management. 
One strategy is to highlight the jobs that are most likely to fail, prior to their execution on the system. Jobs failing during their execution unnecessarily occupy resources which could delay other jobs, adversely affecting the system performance and energy consumption. 
We distinguish between failures due to external factors, such as problems with the computing nodes, networking issues, workload manager downtime (\emph{exogenous} failures)\cite{rojas2019analyzing}, and those due to internal reasons, such as wrongly configured submission scripts and software bugs (\emph{endogenous} failures)\cite{di2019characterizing}. We here focus on the latter category. Forecasting failures due to internal factors a priori would allow to adopt ad-hoc workload management strategies. 


In this paper, we present a Machine Learning (ML) based classification approach to predict endogenous job failures. Our approach is applicable to data that can be collected from a production machine and leverages only the information available at job submission time (hence does not require any instrumentation of the users' code nor any change to standard workload submission workflow). This information might have different formats, and text is among them. To extract more meaningful job information from such textual data, we employ Natural Language Processing (NLP) tools and improve the classification performance of the ML models. To the best of our knowledge, this is the first work that exploits an NLP method  
to represent jobs during classification.
Contrary  to the majority of the past studies which work on random splits of historical data, the proposed methodology can be deployed in an \emph{online} context where jobs are continuously submitted by users to a real production system. 
We demonstrate the validity of our approach on a dataset collected from a production machine Marconi100 hosted at the HPC centre CINECA\footnote{ \url{https://www.hpc.cineca.it/hardware/marconi100}.} in Italy.


\section{Related Work}

In this paper, we restrict the related work to the study of failures in large-scale systems at job/application level. In \cite{Fadishei2009}, the authors analysed workload traces in a grid, showing the correlations between failure characteristics and performance metrics.  Works like \cite{Chen2014,Islam2017} tackled application failure prediction in cloud computing by using recurrent neural networks on resource usage data and performance logs, extracted from Google cluster workload traces. Also in \cite{Yoo2016} the authors relied on the resource usage data of a job to predict its failure, but in the scope of an HPC center. 

These approaches do not take into account the human factors (error in the code, the submission, etc.), which are responsible for many job failures \cite{li2023workload}.
Therefore, the trend is shifting towards the use of data collected from a workload manager to predict failure using job features, as done in \cite{li2023workload,jassas2022,Banjongkan2021}.
In \cite{Banjongkan2021}, the authors use a decision tree algorithm to predict job failure on two HPC workloads.
In \cite{jassas2022}, they survey several ML techniques to perform the same task on a Google cluster workload trace and other two HPC workloads. 
A similar approach is reported in \cite{li2023workload} on 
another 
workload; in addition, they use NLP techniques to assign similar names to similar jobs executed by the same user. All this past work, which are most related to ours, evaluate their approach on random splits of data, which is not realistic because testing could be done on data which is chronologically placed in between the training data traces. Our work differs in two ways:  (i) we propose to use NLP techniques to represent jobs for classification via all the job information available at job submission time, 
(ii) our approach can be deployed in a more realistic \textit{online} context and is thus evaluated on a streaming data, by continuously retraining the classification model on recent (past) data, and testing it on (future) data  which has not been seen.

\section{Background}

In this section, we first present our workload dataset and then the ML models we employ for job failure prediction.

\subsection{M100 Dataset}
\label{M100Workload}

The data used in this study is extracted from the M100 workload \cite{andrea_borghesi_2023_m100} which is the result of more than two years of monitoring on Marconi100, an HPC system hosted at CINECA\footnote{\url{https://www.cineca.it}} in Italy. Marconi100 is a tier-0 supercomputer deployed in production since May 2020 and, at the time of writing, is ranked $24^{th}$ in the top500 list\footnote{\url{https://www.top500.org}}. The cluster is composed of 980 computing nodes, each equipped with two 16-cores IBM POWER9 AC922 processors at 3.1 GHz, four NVIDIA Volta V100 GPUs, and 256 GB RAM. The resources are accessed through eight login nodes, and all the components are connected by a Mellanox Infiniband EDR DragonFly+ 100 Gb/s network infrastructure.  Resources are allocated through job submission to Slurm, 
the workload manager installed in the system. 


M100 contains data ranging from the computing nodes' internal information such as core load, temperature, power consumption, to the system-wide information, including the liquid cooling infrastructure, the air-conditioning system, the power supply units, workload manager statistics, and job-related information. 
For the purposes of our work, we focus on the data which describes the jobs present in the workload by 
features related to their submit-time, run-time and end-time. The first category contains the information available when a job is submitted, such as submission time, requested resources, user information and system state. The second category comprises the information about the job launch, such as waiting time, execution start time, and the actually allocated resources. 
At job termination, the end-time features are collected, e.g., ending time, duration and outcome of the execution.  The full list of job features 
is available at the dataset repository.\footnote{\url{https://gitlab.com/ecs-lab/exadata/-/blob/main/documentation/plugins/job\_table.md}}

One feature related to the execution outcome is the job   
Exit State (ES) label, 
which is assigned to each job by Slurm as an interpretation of the job's Exit Code (EC). This code is formed by a pair of numbers; we consider only the first one, which refers to a system response that reports success, failure, or the reason of an unexpected result from job launch. An EC value of 0 means successful completion, while any EC $\neq 0$ represents an error encountered during execution. Table \ref{ext_state_orig} describes the ES labels assigned to the jobs in our dataset, along with  their distribution. As seen in the table, the dataset is highly unbalanced. This is not surprising, because in a real production machine the failures should be minimized to guarantee correct functioning of the system. 
Nevertheless, 
the percentage of the jobs not successfully completed is more than 20\% (more than 1 out of 6 millions of jobs), representing an important threat to the system performance.

\begin{table}[t!]
\scriptsize
\begin{center}

    \begin{tabular}{|l|l|l|}
    \hline
    \textbf{Name} &  \textbf{Description} & \textbf{\%}\\
    \hline
    Completed & Job completed execution without errors &  79\% \\
    Failed & Job terminated for an unknown reason & 10\% \\
    Cancelled & Job did not start execution due to an error in submission & 8\% \\
    Timeout & Job terminated due to reaching the time limit & 2\% \\
    Out of memory & Job terminated due to more memory access than allocated & 0.6\%\\
    Preempted & A higher-priority job delayed the job execution & 0.1\%\\
    Node fail & Job terminated due to a failure in an allocated node & 0.01\%\\
    \hline
    \end{tabular}
    \end{center}
    \caption{Job ES labels and their distribution in the M100 dataset.}\label{ext_state_orig}
\end{table}

\subsection{Classification and NLP Models}
\label{machine learning models}

We approach the prediction task as a binary classification problem. We exploit supervised and unsupervised techniques for classification, as well as a pre-trained state-of-the-art NLP model to represent jobs during classification. 

As for supervised algorithms, we consider the widely adopted Decision Tree, Random Forest and Logistic Regression. Decision Tree (DT) is a non-parametric method used for classification and regression, which predicts the value of a target variable by learning simple decision rules inferred from the data features. Random Forest (RF) is an ensemble method based on creating a diverse set of DT classifiers by introducing randomness in each DT construction. The prediction of the ensemble is given as the averaged prediction of the individual classifiers. Individual DTs typically exhibit high variance and tend to overfit. The aim of the ensemble method is to remove the error by taking an average of those predictions. Logistic Regression (LR) instead maps the probability of a label given the features of the data. It is usually faster than the other techniques and because of that is one of the most popular classification algorithms.

As an unsupervised algorithm, we employ $k$-Nearest Neighbors (KNN) which is a type of instance-based learning that does not construct a general internal model,
but rather project data points into a $N-$dimensional feature space 
and then consider their distances.
Classification is computed from a simple majority vote of the $k-$nearest neighbors of each data point, where $k$ is a hyperparameter. The $k-$nearest neighbors are computed based on a distance metric, which could be for instance 
Cosine Distance (CD) and Minkowski Distance of order $p$ (MWD$_p$). Given two vectors $X$ and $Y$, representing the data points in the feature space, the distances are 
calculated as $CD(X,Y) = 1 - \cos{\theta} = 1 - \frac{X \cdot Y}{\|X\| \|Y\|}$ and $MWD_p(X,Y) = \sqrt[p]{ \sum_{i = 1}^{n} {\left|X_i - Y_i \right|}^{p}}$
where $\theta$ is the cosine of the angle between the vectors and MWD$_p$ is a generalization of the Euclidean distance. 




Sentence-BERT (SBERT) \cite{reimers2019sentencebert} is a modification of the pre-trained BERT \cite{devlin-etal-2019-bert} language model. BERT is a well-known family of models based on the transformer architecture \cite{vaswani2017attention}, used to give a numeric representation of words (or subwords) that takes into account the context in which these words are used. While BERT works well with classification tasks, it does not work equally well with regression tasks, such as sentence similarity. SBERT produces representations of sentences, not individual words, that are particularly apt for regression tasks. The representation of a string of text produced by SBERT is a fixed-size 384-dimensional floating-point array.

\section{Methodology}
\label{sec:method}

In this section, we describe our methodology to  job failure prediction. The workflow can be divided into two phases: (i) data preparation and (ii) job failure prediction. 

\subsection{\textbf{Data preparation}}
\label{data prep}



To train and test our classifiers, we consider a part of the dataset\footnote{\url{https://doi.org/10.5281/zenodo.7588815}} and use only the data collected between May 2020 and October 2020. The reason is that this is the only period where the dataset contains information on the requested resources and the job EC, which we need for our  prediction task.  
We collect the job data in a data frame and then prepare it for model training and inference. 

\paragraph{\textbf{Feature selection}}
\label{job_structure}

In order to describe the characteristics of a job in a classification task, we need to associate it with certain features. We focus only on job submit-time features, as we want to compute a prediction before job allocation. The features available in the dataset are listed in Table \ref{job_features} along with their description. Jobs submitted by the same user and close in time tend to be similar because in a production HPC, users often submit jobs in batches referring to similar experiments and jobs in the same batch tend to have similar names and command. Thus, we believe that all these features are useful for our purposes. We note that user name and similar private data are omitted in the public dataset. However, CINECA granted us access under a non-disclosure agreement.   


\begin{table}[t!]
\scriptsize
\begin{center}
\resizebox{0.9\textwidth}{!}{
    \begin{tabular}{|l|l|l|}
    \hline
    \textbf{Name} &  \textbf{Description} & \textbf{Type}\\
    \hline
    Name & Job name assigned by the user & String \\ 
    Command & Command executed to submit the job & String \\ 
    Account & Account to be charged for job execution & String \\
    User id & ID of the user submitting the job & Integer \\
    Dependency & Jobs to wait for completion before execution & String \\
    Group id & Group of the user submitting the job & Integer  \\
    Requested nodes & Specific nodes requested  & List[String] \\
    Num tasks per socket & Number of tasks to invoke on each socket & Integer \\
    Partition & Name of the assigned partition  & String \\
    Time limit & Maximum allowed run time in minutes or infinite & Integer \\ 
    Qos & requested quality of service & String \\
    Num cpu & Number of rquested CPUs   & Integer \\ 
    Num nodes & Number of requested nodes   & Integer  \\
    Num gpus & Number of requested GPUs & Integer \\ 
    Submit time & Time of job submission & Timestamp \\
    \hline
    \end{tabular}}
\end{center}
\caption{Job features description.}\label{job_features}
\end{table}


\paragraph{\textbf{Job exit state labels}}
\label{job exit state labels}

For the training data, we need to assign a label to each job, indicating whether it has failed or not. In Section \ref{M100Workload}, we presented the job ES labels  as they are present in the dataset, which are assigned by Slurm based on job EC. According to the Slurm official documentation, the labels  assigned by the scheduler may not be coherent with the actual EC, due to lack of proper synchronization between the signal emitted by the job exit and the data collected in the database. We therefore inspect the data and identify any possible discrepancy, e.g., a job with an ES label \textit{completed} and an EC $\neq 0$. Our analysis reveals that more than 70K jobs labelled differently than \textit{completed} have an EC value of 0. This is confirmed by the difference between the percentage of the completed jobs (83\%) and the jobs having an EC of 0 (89\%).
As a consequence, we discard the original labels and create new labels based on the job EC.  

\begin{figure}[t!]
    \centering
    \includegraphics[width=0.65\textwidth, height=2in]
    %
    {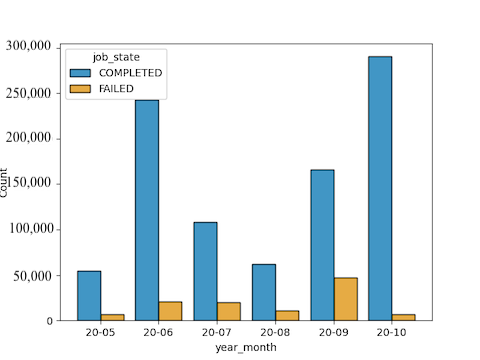}
\captionof{figure}{Job ES label distribution throughout the months in the final dataset.}
\label{figdistmonth}
\end{figure}

Despite the discrepancy between the original ES labels and EC, the highly unbalanced nature of the entire dataset (see Sec.~\ref{M100Workload}) is observed also in the subset data we use in this study. In particular, while the percentage of jobs with EC $= 1$ is 9\%,  the percentage with EC $> 1$ is 2\%. We therefore group all types of failures under the same category; discriminating among different fail modes is outside the scope of this work. Moreover, we are interested in failure caused by the workload itself, so we remove from the dataset all the jobs originally labelled as \textit{cancelled} (failure due to user) and \textit{node fail} (failure due to hardware). Eventually, we re-label the remaining data according to the following policy: for every job, we assign an ES label of \textit{completed} if its EC is 0, \textit{failed} otherwise.
The final dataset after the relabelling is composed of 924,252 (89\%) completed  and 113,027 (11\%) failed jobs. The distribution of the labels, throughout the months, is reported in Figure \ref{figdistmonth}. We can observe that imbalance between the two classes of jobs appears in all the months, while the ratio between them changes considerably, showing that the workload is highly variable across time.

\subsection{Job failure prediction}

\paragraph{\textbf{Feature encoding}}

\label{feat_encoding}

In order to compute a prediction for a job, we need to represent it suitably to feed into the classification models presented in Section \ref{machine learning models}.
We achieve that by relying on job feature values, and we propose two different ways to encode them. 
In the first (INT), we assign an integer to the values which are not numerical, i.e. \textit{name, command, account, dependency, requested nodes, partition, qos, submit time}, while setting all the missing values in the other fields (\textit{num tasks per socket, time limit}) to a default value of 0.  In the second encoding (SB), we first concatenate all the feature values into a comma divided string, e.g. \textit{job1, run\_job1.sh, [1, 10], 2020-10-01 15:30:00, account\_1, partition\_1, 0, normal, 4, 100, 2, etc}. Then we encode the string with SBERT, obtaining a 
384-dimensional floating-point array.

We believe that with SBERT we can extract more fine-grained insights about job features expressed in natural language (e.g. \textit{name, command, account}). This is because SBERT is designed to result in similar encodings with sequences with semantically similar contents. As we discussed in Section \ref{job_structure}, jobs with similar names and command could belong to the same submission batch running similar operations. Therefore, features like \textit{submit\_time, name, account, command} could reveal important patterns on the nature of the job and its workload. This is hard to recognize with the INT encoding, since similar natural language values will be mapped to different integer values, while they would have similar representation in SB, due to semantic similarity.


\paragraph{\textbf{Classifier training and testing}}

In our prediction task, it would not be realistic to do inference on a job by learning from the data of the future jobs  submitted at a later time. We thus create the  training and test sets by considering the timeline of the job data, keeping in the training set the data that comes before in chronological order  the data of the test set. 

We identify two settings in which a classifier can be trained and tested on a dataset. The first is the \textit{offline} setting, where we consider the job data as a whole, train the model once on one portion of it, and test it using the data of the other portion in chronological order. To do this, we sort the jobs based on their submission time, split them into two, use the first split preceding in time as the training set, and the other as the test set. 

The second setting, which we refer to as \textit{online},  is  more suitable to our context. We treat the job data as live and streaming in time, retrain the model periodically on a fixed size of recent data, and test it on future data that comes later (but near) in time. As we discussed in Section \ref{job_structure}, the workload of an HPC system can be very similar in a short period, while may vary  in the long term. As our experimental results confirm, a model trained once on data which slowly gets further in time to the test data 
could classify poorly compared to a model which is retrained continuously on data closer in time to the test data.

In the \textit{online} setting, we use the time information provided by the \textit{submit\_time, start\_time and end\_time} features in order to simulate 
job submission and execution on a machine, and add the \textit{day} feature as the submission date by extracting it from \textit{submit\_time}. 
We consider as the first training set all the jobs that  were submitted in the first $\alpha$ days and not finished after the date of the first test set. Starting from the submission time of the first job not present in the first training set, we divide the data in batches in chronological order, where each batch contains the jobs submitted in the next $\omega$ days. We then iterate over each batch, considering it as a new test set. At every iteration, the training set is updated with the data of the last $\alpha$ days and the supervised models are retrained.  
With the unsupervised models, no actual re-training takes place, however the training set is extended for each new job in the test set  with the jobs that finished before the submission time of the new job (with negligible overhead).


\section{Experimental Study}

In this section, we report our experimental study and discuss our results. 
\paragraph{\textbf{Experimental setting}}

All the experiments are conducted on a node of a small cluster
equipped with two Marvell TX2 CPUs with 32 cores and 256 GB of RAM.  No accelerator, such as GPU, is used in the experiments. 

The classification algorithms are implemented with \textit{scikit-learn} Python library. The sequence encoder model is provided by the \textit{sentence transformers} library\footnote{\url{https://www.sbert.net}}, while the weights for SBERT are pulled from huggingface.\footnote{\url{https://huggingface.co}} We use the pre-trained model \textit{all-MiniLM-L6-v2}\footnote{\url{https://huggingface.co/sentence-transformers/all-MiniLM-L12-v2}}, since it is the best trade-off between prediction performance and speed \cite{reimers2019sentencebert}.
All the models are instantiated with the default setting provided by the library. 

We set the hyperparameters as follows after an initial empirical evaluation.  We use MWD of order $p=2$ and set $k = 5$ in the KNN algorithm. As discussed in Section \ref{sec:method}, the testing period strictly follows the training period. For the offline setting, we take the first 70\% of the data as the training set and the remaining 30\% as the test set. For the online, we fix the training interval $\alpha$ to 30 days, based on the trade-off between prediction performance and training/inference time. The time-span of data in each test set is $\omega = 1$ day. 
The implementation 
is available in a GitHub repository.\footnote{\url{https://github.com/francescoantici/job-failure-predictor/}}

The results are reported in Tables \ref{results_global_seq} and \ref{results_global_online}, where we distinguish between the job feature encodings (INT and SB), the supervised algorithms (DT, LR, RF), and the distance metrics of the KNN algorithm (CD and MWD). Each classification algorithm is evaluated using the two feature encodings and are compared with two simple baselines, namely majority and random. Both baselines ignore the input feature values. The majority returns the most frequent label observed in the training data, while the random generates predictions uniformly from the list of unique labels, so each class has equal probability. The results reported in Table \ref{results_global_online} are averaged over 5 months between June 2020 and October 2020.

\paragraph{\textbf{Results}}

\begin{table}[t!]
\scriptsize
\begin{center}
    \begin{tabular}{|l|l|l|l||l|l|l||l|l|l|}
    \hline
    Model & T F1$_m$ & T Prec$_m$ & T Rec$_m$ & C F1 & C Prec & C Rec & F F1 & F Prec & F Rec\\
     
    \hline
    Supervised  & & & & & & & & & \\
    INT+DT & 0.30 & 0.50 & 0.48 & 0.55 & 0.96 & 0.38 & 0.06 & 0.03 & \textbf{0.57}\\
    INT+LR & 0.54 & 0.62 & 0.53 & \textbf{0.98} & 0.97 & 0.99 & 0.10 & 0.26 & 0.06\\
    INT+RF & \textbf{0.71} & \textbf{0.72} & \textbf{0.69} & \textbf{0.98} & \textbf{0.98} & 0.98 & \textbf{0.43} & \textbf{0.47} & 0.39 \\
    SB+DT &  0.38 & 0.50 & 0.50 & 0.70 & 0.97 & 0.55 & 0.06 & 0.03 & 0.45 \\
    SB+LR & 0.66 & 0.70 & 0.63 & \textbf{0.98} & \textbf{0.98} & 0.99 &  0.34 & 0.43 & 0.28 \\
    SB+RF & 0.55 & 0.54 & 0.61 & 0.95 & 0.97 & 0.92 & 0.16 & 0.11 & 0.30 \\
    \hline
    Unsupervised  & & & & & & & & & \\
    INT+CD & 0.52 & 0.52 & 0.58 & 0.92 & 0.97 & 0.87 & 0.11 & 0.07 & 0.28\\
    INT+MWD & 0.39 & 0.50 & 0.50 & 0.72 & 0.97 & 0.58 & 0.06 & 0.03 & 0.42 \\

    SB+CD &  0.42 & 0.50 & 0.52 & 0.76 & 0.97 & 0.63 & 0.07 & 0.04 & 0.42\\
    
    
    SB+MWD & 0.42 & 0.50 & 0.52 & 0.76 & 0.97 & 0.63 & 0.07 & 0.04 & 0.42  \\
    
    \hline
    Majority & 0.49 & 0.50 & 0.48 & \textbf{0.98} & 0.97 & \textbf{1.00} & 0.00 & 0.00 & 0.00 \\
    Random  & 0.36 & 0.50 & 0.50 & 0.66 & 0.97 & 0.50 & 0.06 & 0.03 & 0.49  \\
    \hline
    \end{tabular}
    \end{center}
    \caption{Results in the offline setting, for both classes (T), completed class (C) and failed class (F) using precision (Prec), f1 and recall (Rec). 
    In (T),  we consider the macro averaged metrics (F1$_m$, Prec$_m$, Rec$_m$). The model name is composed of the feature encoding and the classification algorithm/distance metric. Best results are highlighted in bold. }
    \label{results_global_seq}
\end{table}


We evaluate our models with metrics typically used for classification tasks, namely f1, precision and recall.
Table \ref{results_global_seq} reports the results of the offline setting.  The model that gives the best results overall is INT+RF. It achieves a f1 score of 71\% and is very good at classifying the completed jobs, as the f1 score computed over such jobs is 98\%. The prediction of the failures is somewhat harder, with a f1 score of 43\%. 

Overall, we observe that the supervised techniques perform better, but all the models struggle with the classification of the failed jobs, as most of them (with the exception of INT+DT) have lower recall than the random baseline in the failed class.
Conversely, the classification of completed jobs is much easier, with the precision being $\geq$ 96\%; this is probably due to the imbalance in the dataset (completed jobs are more abundant). This is compounded with the proportion between the completed and failed jobs varying significantly across different periods, as shown in Figure \ref{figdistmonth}. 
Thus, with the offline setting, the model has a high risk of overfitting on the completed job examples (being more numerous) and of spectacularly underperforming when tested on jobs that fail.

This behaviour can be mitigated by retraining the models to adapt them to the workload and the class distribution shift over time. 
Indeed, Table \ref{results_global_online} shows the results of the online setting, with notable improvements in the classification of the failed jobs. 
The SB encoding coupled with the clustering classifier using the Minkowski distance (SB+MWD) yields the best results overall, suggesting that properly extracting meaningful job information from textual data is beneficial.
In terms of the f1 score, SB+MWD  reaches 70\%, outperforming all the supervised models, which arrive to a maximum of 64\% with SB+RF and INT+RF.

\begin{table}[t!]
\scriptsize
\begin{center}
    \begin{tabular}{|l|l|l|l||l|l|l||l|l|l||l|}
    \hline
    Model & T F1$_m$ & T Prec$_m$ & T Rec$_m$ & C F1 & C Prec & C Rec & F F1 & F Prec & F Rec & Time \\
    \hline
    Supervised  & & & & & & & & & &\\
    INT+DT  & 0.60 & 0.64 & 0.63 & 0.80 & 0.84 & 0.79 & 0.41 & 0.44 & 0.46 & 1.27 + 0.005 \\
    INT+LR  & 0.46 & 0.53 & 0.51 & 0.85 & 0.79 & 0.95 & 0.06 & 0.26 & 0.06 & 78 + 0.3\\
    INT+RF & 0.64 & 0.69 & 0.64 &  0.84 & 0.84 & 0.87 &  0.43 & 0.54 & 0.41 & 25 + 0.12 \\
    
    SB+DT & 0.61 & 0.62 & 0.63 &  0.80 & 0.84 & 0.78 & 0.41 & 0.39 & 0.47 & 455 + 0.09 \\
    SB+LR  & 0.60 & 0.66 & 0.60 & 0.85 & 0.82 & 0.89 & 0.34 & 0.50 & 0.30 & 84 + 0.4\\
    SB+RF  & 0.64 & 0.70 & 0.63 & 0.86 & 0.83 & 0.91 & 0.41 & \textbf{0.57} & 0.35  & 922 + 0.4\\
    \hline
    Unsupervised & & & & & & & & & &\\
    INT+CD  & 0.68 & 0.69 & 0.69 & 0.84 & 0.86 & 0.82 & 0.52 & 0.52 & 0.56 & \textbf{N.A. + 0.3}\\
    INT+MWD  & 0.68 & 0.69 & 0.69  & 0.84 & \textbf{0.87} & 0.83 & 0.52 & 0.51 & 0.55 & \textbf{N.A. + 0.3}\\
    
    SB+CD & 0.69  & 0.70 & 0.71 & 0.84 & \textbf{0.87} & 0.83 & \textbf{0.54} & 0.54 & \textbf{0.59} & N.A. + 0.7\\
    SB+MWD  & \textbf{0.70} & \textbf{0.71} & \textbf{0.71} & 0.85 & \textbf{0.87} & 0.83 & \textbf{0.54} & 0.54 & \textbf{0.59} & N.A + 0.7 \\
    \hline
    Majority  & 0.44 & 0.40 & 0.50 & \textbf{0.87} & 0.79 & \textbf{1.00} & 0.00 & 0.00 & 0.00 & N.A. \\
    Random  & 0.44 & 0.50 & 0.50 & 0.61 & 0.79 & 0.5 & 0.28 & 0.21 & 0.5 & N.A. \\
    \hline
    \end{tabular}
    \end{center}
     \caption{
     Results in the online setting, presented similarly to Table \ref{results_global_seq}.  The time (in sec) is the avg. training time per day and the avg. inference time per job (including the SB encoding time where applicable -- “N.A.” indicates the cases where SB is not applicable). }
    \label{results_global_online}
\end{table}

The classification of the completed jobs is good for all the models and their f1 scores are always above the 80\%; the clustering methods have the highest precision (87\%), while SB+RF has better recall (91\% with respect to 83\%).
There is some minor drop in performance in the completed class compared to the offline setting (less overfitting), but the results are still solid. In the failed class, the clustering methods (SB+CD, SB+MWD) obtain a f1 score of 54\% outperforming all the supervised algorithms. We observe a significant improvement with respect to the offline setting. Indeed, the best f1 score obtained  over failed jobs in the offline setting  (INT+RF) is increased by 20\% by the best model in the online setting (SB+MWD and SB+CD); clearly, retraining the models helps to classify job failures. 

As can be observed in both tables, the use of the SB encoding has a marginal impact with the supervised models, while the training time increases significantly in the online context (e.g., the training time of INT+RF is 25 seconds, while SB+RF requires 922 seconds). The increase in training time is not surprising, as the extraction of the text features through NLP involves the usage of a computationally hungry DN. We note, however, that the inference time remains very small and this is the operation that needs to be performed in real time without affecting the machine's normal workload (the retraining can be scheduled in less busy periods). On the other hand, in the case of the unsupervised models, SB improves the performance by 1-2\% in almost every metric while no training time is incurred and the inference time always remains under a second. As we discussed in Section \ref{sec:method}, with these models retraining is simply extending the training set (with negligible overhead) and classifying a new job requires a simple inference step (i.e., the new job is compared with those in the training set, projected in the feature space).


\section{Conclusions and Future Work}



We presented an ML-based classification approach to predict endogenous job failures in HPC systems, using only the information available at job submission time. The methodology can be deployed in an \textit{online} context where jobs are continuously submitted by users to a real production system. 
We thoroughly validated our approach with a two-fold battery of test using supervised and unsupervised learning algorithms.
In the first, we considered an \textit{offline} setting and split the job data in time-consecutive sets for training and testing. We showed that in this setting the models poorly classify the failed jobs -- which is what we are more interested in -- while they are pretty accurate in predicting the completion.
We then deployed our approach \textit{online}, where we treated the job data as live and streaming in time, retrained the model periodically on recent (past) data, and tested it on (future) data that comes later (but near) in time. We observed an improvement in prediction accuracy by the use of this setting, especially in predicting the job failures. We also showed that an unsupervised technique like KNN is more suitable in the online setting, and the use of an NLP-based encoding to represent job features improves the classification accuracy.

Our contribution can be seamlessly integrated into the existing operational data analytic frameworks deployed in modern systems. The marginal overhead increase 
is not worrying, as adopting hardware accelerators (GPUs, TPUs, etc.) or deploying the models to scalable architectures will make the inference time almost negligible. In future work, we want to study continuous learning techniques and investigate different retraining strategies.
We also plan to take into account the uncertainty of the ML models and investigate  policies to handle jobs with high failure risk (in accordance to the Service-Level-Agreements (SLAs) between the HPC provider and the users).  For instance, the workload deemed to be at high risk of failure can be postponed, and the user can be asked to revise the job submission. As another example, the high-risk workload (according to failure classification) can be directly discarded if the confidence of the classifier surpasses a threshold defined by the SLA. The user can then be encouraged to resubmit, which can be treated as higher priority not to incur in additional delays.


%
%
%
\bibliographystyle{splncs04}
\bibliography{bibtext.bib}
%




\end{document}